\documentclass{article}

\usepackage{arxiv}

\usepackage{amsthm}
\usepackage{amssymb}
\usepackage{mathtools}
\usepackage{mathpartir}
\usepackage{stmaryrd}
\usepackage{leftindex}
\usepackage{wasysym}
\usepackage{esvect}
\usepackage{wasysym}

\usepackage[utf8]{inputenc} 
\usepackage[T1]{fontenc}    
\usepackage{hyperref}       
\usepackage{url}            
\usepackage{booktabs}       
\usepackage{amsfonts}       
\usepackage{nicefrac}       
\usepackage{microtype}      
\usepackage{cleveref}       
\usepackage{lipsum}         
\usepackage{graphicx}
\usepackage{natbib}
\usepackage{doi}

\title{Hunting DeFi Vulnerabilities via Context-Sensitive Concolic Verification\thanks{This is the preprint version of the conference paper presented in \textit{International Conference on Software Engineering, 2024}.}}

\date{}

\newif\ifuniqueAffiliation
\uniqueAffiliationtrue

\ifuniqueAffiliation 
\author{ \href{https://orcid.org/0000-0002-6996-9333}{\includegraphics[scale=0.06]{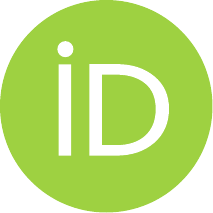}\hspace{1mm}Yepeng Ding}\\
	The University of Tokyo\\
        Tokyo, Japan\\
	\texttt{yepengd@acm.org} \\
	\And
	\href{https://orcid.org/0000-0002-3565-3410}{\includegraphics[scale=0.06]{orcid.pdf}\hspace{1mm}Arthur Gervais} \\
	University College London\\
	London, United Kingdom\\
	\texttt{arthur@gervais.cc} \\
	\And
	\href{https://orcid.org/0000-0002-6339-3134}{\includegraphics[scale=0.06]{orcid.pdf}\hspace{1mm}Roger Wattenhofer} \\
	ETH Zurich\\
	Zurich, Switzerland\\
	\texttt{wattenhofer@ethz.ch} \\
 	\And
	\href{https://orcid.org/0000-0002-2891-3835}{\includegraphics[scale=0.06]{orcid.pdf}\hspace{1mm}Hiroyuki Sato} \\
	The University of Tokyo\\
	Tokyo, Japan\\
	\texttt{schuko@satolab.itc.u-tokyo.ac.jp} \\
}
\else
\usepackage{authblk}

\newbox{\orcid}\sbox{\orcid}{\includegraphics[scale=0.06]{orcid.pdf}} 
\author[1]{%
	\href{https://orcid.org/0000-0000-0000-0000}{\usebox{\orcid}\hspace{1mm}David S.~Hippocampus\thanks{\texttt{hippo@cs.cranberry-lemon.edu}}}%
}
\author[1,2]{%
	\href{https://orcid.org/0000-0000-0000-0000}{\usebox{\orcid}\hspace{1mm}Elias D.~Striatum\thanks{\texttt{stariate@ee.mount-sheikh.edu}}}%
}
\affil[1]{Department of Computer Science, Cranberry-Lemon University, Pittsburgh, PA 15213}
\affil[2]{Department of Electrical Engineering, Mount-Sheikh University, Santa Narimana, Levand}
\fi

\renewcommand{\shorttitle}{Hunting DeFi Vulnerabilities via Context-Sensitive Concolic Verification}

\begin{document}
\maketitle

\begin{abstract}
Decentralized finance (DeFi) is revolutionizing the traditional centralized finance paradigm with its attractive features such as high availability, transparency, and tamper-proofing. However, attacks targeting DeFi services have severely damaged the DeFi market, as evidenced by our investigation of 80 real-world DeFi incidents from 2017 to 2022. Existing methods, based on symbolic execution, model checking, semantic analysis, and fuzzing, fall short in identifying the most DeFi vulnerability types. To address the deficiency, we propose Context-Sensitive Concolic Verification (CSCV), a method of automating the DeFi vulnerability finding based on user-defined properties formulated in temporal logic. CSCV builds and optimizes contexts to guide verification processes that dynamically construct context-carrying transition systems in tandem with concolic executions. Furthermore, we demonstrate the effectiveness of CSCV through experiments on real-world DeFi services and qualitative comparison. The experiment results show that our CSCV prototype successfully detects 76.25\% of the vulnerabilities from the investigated incidents with an average time of 253.06 seconds.
\end{abstract}


\section{Introduction}

Decentralized Finance (DeFi) has been marred by significant security challenges. In our investigation, 80 real-world DeFi incidents that occurred between November 2017 and December 2022 have proved to be considerable threats to the stability of the DeFi market, resulting in financial damages ranging from 2,400 to 600 million dollars. We classify their underlying vulnerabilities into six types based on their root causes and sort their severity regarding the average loss in US dollars per incident, as detailed in Table~\ref{tab:defi_vulnerabilities}.

\begin{table}[h]
\centering
  \caption{Investigated Typical DeFi Vulnerabilities}
  \label{tab:defi_vulnerabilities}
  \begin{tabular}{llcc}
    \toprule
    \textbf{Root Cause} & \textbf{Description} & \textbf{Total} & \textbf{Loss}\\
    \midrule
    BF & Business Logic Flaw & 23 & \$1.4B \\
    RE & Reentrancy & 9 & \$147M \\
    PM & Price Oracle Manipulation & 20 & \$92M \\
    IV & Insufficient Validation & 10 & \$12M \\
    AF & Access Control Flaw & 13 & \$14M \\
    UE & Unexpected External Call & 5 & \$3M \\
  \bottomrule
\end{tabular}
\end{table}

Unfortunately, the three most severe root causes, accounting for the largest financial losses, pose a formidable challenge for existing methods. \underline{BF} (e.g., the Eleven Finance hack in 2021) and \underline{RE} variants (e.g., the Rari Capital hack in 2022), which exploit conformance issues between requirement specifications and implementations, render highly automated methods (\cite{frank_ethbmc_2020, nguyen_sfuzz_2020}) ineffective, while other methods (\cite{choi_smartian_2021,so_smartest_2021}) struggle with scalability when addressing these non-patterned vulnerabilities. Besides, \underline{PM} vulnerabilities offer attackers lucrative arbitrage opportunities with minimal attack costs via flash loans, yet current methods based on control flow (\cite{frank_ethbmc_2020}), data flow analysis (\cite{choi_smartian_2021}), symbolic execution (\cite{so_smartest_2021}), and fuzzing (\cite{nguyen_sfuzz_2020}) fall short in identifying these plausibly normal financial behaviors.

Motivated by addressing the challenge, we propose \underline{C}ontext-\underline{S}ensitive \underline{C}oncolic \underline{V}erification (CSCV) to automatically find all classified types of DeFi vulnerabilities by user-defined temporal properties. Concolic verification distinguishes itself from standard concolic execution by leveraging formal verification, effectively synergizing concolic execution (\cite{baldoni_survey_2018}) with model checking (\cite{clarke2018handbook}).

\section{CSCV in a Nutshell}

\begin{figure}[h]
  \centering
  \includegraphics[width=0.8\linewidth]{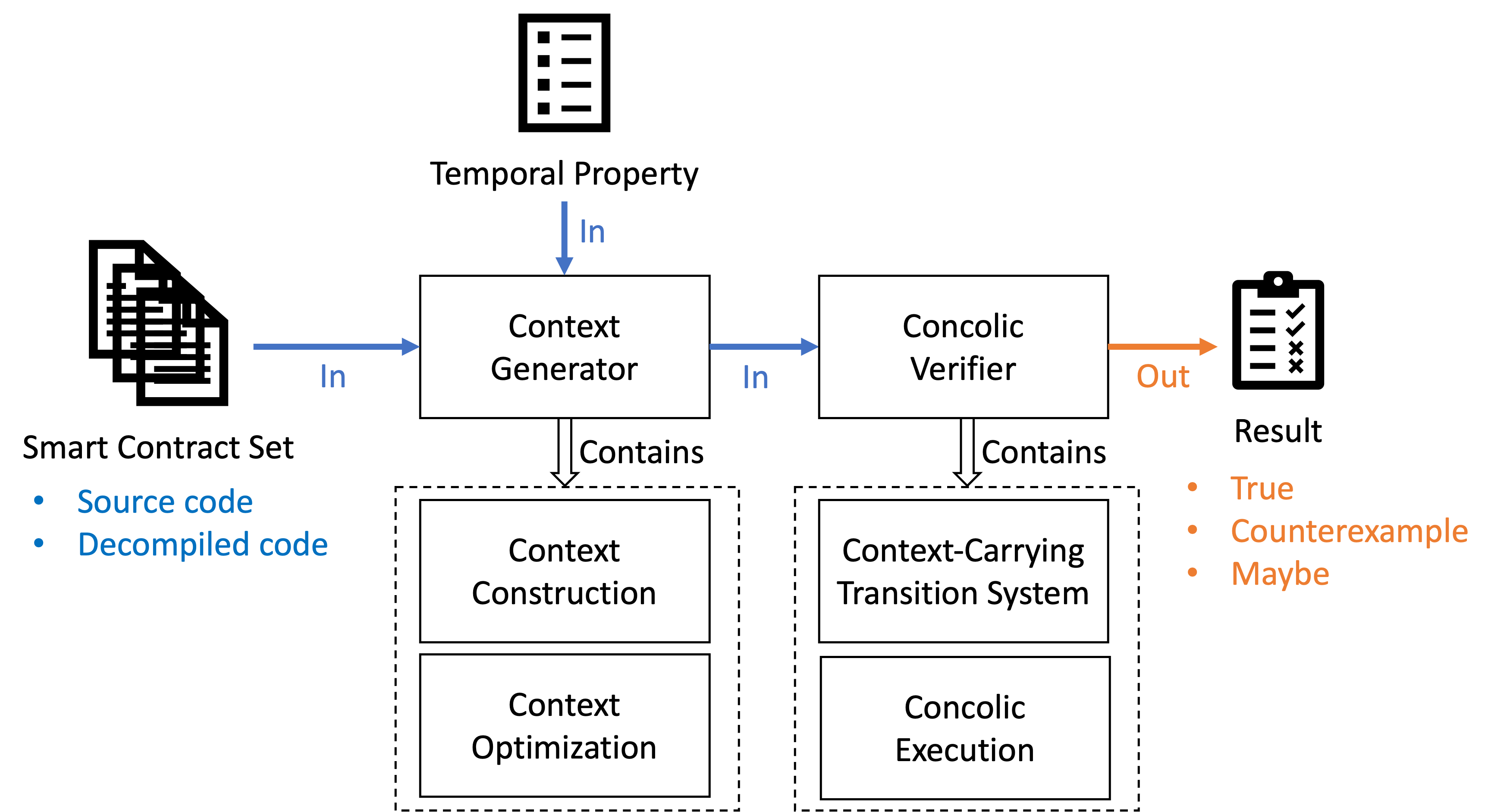}
  \caption{Overview of CSCV.}
  \label{fig:overview}
\end{figure}

In concolic verification of a temporal property $\phi$ specifying DeFi service $\mathfrak{P}$, finding a vulnerability is framed as searching an attack vector, an alternating sequence comprising global states $\vv{S}$ and invocation of functions selected from set $F$ of external write functions, such that $\exists s \in \vv{S}: s \not\models \phi$. However, the brute-force enumeration of all possible combinations of state variables and functions to identify an attack vector is computationally infeasible, given the enormous number of potential permutations. Therefore, we formulate CSCV by introducing contexts to guide concolic verification processes that dynamically construct context-carrying transition systems in tandem with concolic execution during transitions to find attack vectors. The overview of CSCV is visually shown in Figure~\ref{fig:overview}.

\paragraph{Context Construction}
The context generator processes $\mathfrak{P}$ along with $\phi$ to construct a context that encapsulates an \textbf{evaluation function} of state variables at a specific block, and a \textbf{relevance function} hierarchizing $F$ for each $f \in F$. The context construction is governed by a property-based algorithm that filters out extraneous information regarding $\phi$. The algorithm formulates the smallest evaluation function by analyzing the state variable dependency and shapes the relevance function by ranking the self-excluded $F$ based on the number of commonly shared state variables.

\paragraph{Context Optimization}
\textbf{Property spatialization} mitigates the state explosion problem in temporal property verification by encoding temporal formulas into non-temporal assertions that are coded as preconditions and postconditions into respective functions in $F$. \textbf{Function constantization} simplifies the execution logic by substituting free nullary read function calls with constants based on the dependency analysis regarding modified state variables. \textbf{Heuristic identification} accelerates concolic verification processes by pinpointing heuristics obtained from properties, business logic, and historical incidents in context elements.

\paragraph{Concolic Verification}

Concolic verification dynamically constructs a context-carrying transition system that steers concolic executions. Each conditional transition corresponds to an external write function invocation, where conditions are derived from the preconditions encoded from $\phi$. The nondeterminism of transition selection is resolved by the relevance function in contexts. Each transition also initiates a concolic execution following the shortest execution path from the initial state to the current state. If a postcondition is violated, the concolic verifier reports a counterexample and terminates. In cases when the completeness threshold is guaranteed, the concolic verifier confirms that $\phi$ holds. Otherwise, the concolic verifier terminates within a predefined time frame or diameter.

\section{Evaluation}

We mechanized CSCV into a prototype primarily in Java and backed by the Z3 solver for experiments.

\paragraph{Effectiveness}

Our experiments are designed to selectively utilize a specific proportion (0\%, 25\%, 50\%, and 75\%) of heuristics, randomly chosen from an established heuristic base for each investigated DeFi vulnerability in Table~\ref{tab:defi_vulnerabilities}. The experiment results show that our prototype successfully identified 38 (47.50\% of the total) vulnerabilities and 432 attack vectors, including all six classified vulnerability types, even with 0\% heuristics. Moreover, with 75\% heuristics, our prototype identified 61 vulnerabilities (76.25\% of the total) and 1,498 attack vectors, including 20.96\% of previously unknown attack vectors, with an average time of 253.06 seconds.

\paragraph{Comparison}
Our qualitative evaluation offers a comparison between the CSCV methodology and existing methods across a set of criteria, including: vulnerable function path finding (PF), malicious assignment generation (AG), code-level property specification (CP), protocol-level property specification (PP), cross-contract analysis (CC), and DeFi-focus analysis (DF). The evaluation results, detailed in Table~\ref{tab:feature_comparison}, demonstrate the potential of CSCV in addressing the challenge of effectively identifying various DeFi vulnerabilities.

\begin{table}[h]
\centering
  \caption{Feature comparison with existing methods.}
  \label{tab:feature_comparison}
  \begin{tabular}{lcccccc}
    \toprule
    \textbf{Method} & \textbf{PF} & \textbf{AG} & \textbf{CP} & \textbf{PP} & \textbf{CC} & \textbf{DF} \\
    \midrule
    sFuzz (\cite{nguyen_sfuzz_2020}) & \LEFTcircle & \CIRCLE & \Circle & \Circle & \Circle & \Circle \\
    ETHBMC (\cite{frank_ethbmc_2020}) & \Circle & \CIRCLE & \LEFTcircle & \LEFTcircle & \LEFTcircle & \Circle \\
    Smartian (\cite{choi_smartian_2021}) & \Circle & \CIRCLE & \CIRCLE & \Circle & \Circle & \Circle \\
    SmarTest (\cite{so_smartest_2021}) & \Circle & \CIRCLE & \CIRCLE & \Circle & \Circle & \Circle \\
    \hline
    CSCV & \CIRCLE & \CIRCLE & \CIRCLE & \CIRCLE & \CIRCLE & \CIRCLE \\
  \bottomrule
  \multicolumn{7}{l}{\CIRCLE: full support, \LEFTcircle: partial support, \Circle: no support.}
  \end{tabular}
\end{table}

\bibliographystyle{unsrtnat}
\bibliography{references}  






\end{document}